\documentclass[]{JHEP3}
\usepackage{epsfig,multicol,bbm}

\newcommand\fverb{\setbox\fverbbox=\hbox\bgroup\verb}
\newcommand\fverbdo{\egroup\medskip\noindent%
                        \fbox{\unhbox\fverbbox}\ }
\newcommand\fverbit{\egroup\item[\fbox{\unhbox\fverbbox}]}
\newbox\fverbbox

\title{Possibility of Determining $\tau$ Lepton
Electromagnetic  Moments in ${\gamma\gamma \to \tau^{+}\tau^{-}}$
Process  at the CERN-LHC}

\author{S. Ata\u{g}\\
Department of Physics, Faculty of Sciences,
Ankara University, 06100 Tandogan, Ankara, Turkey\\
E-mail:\email{atag@science.ankara.edu.tr}}

\author{A.A. Billur \\
\thanks{Also at Department of Physics, Faculty of Sciences,
Ankara University, 06100 Tandogan, Ankara, Turkey}
Department of Physics, Cumhuriyet University,
58140, Sivas, Turkey\\
E-mail:\email{abillur@science.ankara.edu.tr}}

\abstract{Potential of the LHC  to determine  
the electromagnetic couplings 
of the $\tau$ lepton is discussed via the process 
${\gamma\gamma \to \tau^{+}\tau}^{-}$. 
Highly improved constraints of the anomalous magnetic  
and electric dipole moments  
have been obtained  compared to the LEP  sensitivity.}


\begin{document}

\section{Introduction}

The magnetic moment of the electron which is responsible for 
the interaction with the magnetic field in the Born approximation
can be written in the standard form 

\begin{eqnarray}
\vec{\mu}=g(\frac{e\hbar}{2mc})\vec{s}
\end{eqnarray}
where $\vec{s}$, e and m  is the  spin, electric charge  
and mass of the electron. The cofficient $g$ is 
called the Lande g-factor or gyromagnetic factor.
Standard prediction of the Dirac equation gives $g=2$. 
Deviation from the Dirac value 

\begin{eqnarray}
a_{e}=(g-2)/2
\end{eqnarray}
is known as the anomalous magnetic moment. The first result 
for the anomalous magnetic moment of the electron was calculated 
from Quantum Electrodynamics (QED) using  radiative 
corrections by Schwinger in 1948 as 
$a_{e}=\frac{\alpha}{2\pi}$ \cite{schwinger}.   
From that time, physicists have improved successively 
the accuracy of the this quantity
both in the theoretical and experimental point of views.
These works have provided the stringent tests of QED 
and have lead to the precise determination of 
the fine structure constant $\alpha$ based on the fact that 
$a_{e}$ is insensitive to the weak and strong interactions.
Similar studies have been done for muons. Since the higher 
loop corrections are mass dependent,
the $a_{\mu}$ is expected to include  weak and hadronic 
contributions. This offers a sensitivity to new physics
by a relative enhancement 
factor of $(m_{\mu}/m_{e})^{2}\sim 4\times 10^{4}$ than 
to the case of $a_{e}$. Several detailed Standard Model 
tests have been done using the accurate value of the 
anomalous magnetic moment of the muon \cite{brown}.
Anomalous magnetic moment $a_{\tau}$  of $\tau$ lepton would 
be much better to constrain the new physics due to 
its large mass. However, spin precession experiment
is not convenient to make a direct measurement for 
$a_{\tau}$ at present because of its short lifetime. 
So we need collider experiments with high accuracy
to produce $\tau$ lepton.
Latest  QED contribution to the anomalous 
magnetic moment $a_{\tau}$ from higher loop corrections
is given by the following theoretical result \cite{passera}

\begin{eqnarray}
a^{QED}_{\tau}=117324\times 10^{-8}
\end{eqnarray} 
with the uncertainty $2\times 10^{-8}$.
The experimental limits at 95\% CL  were obtained by L3 and OPAL 
collaborations in radiative $Z\to \tau\tau\gamma$ events 
at LEP \cite{l3,opal} 
\begin{eqnarray}
-0.052 < a_{\tau} < 0.058 \,\,\, \mbox{(L3)}\\
-0.068 < a_{\tau} < 0.065 \,\,\, \mbox{(OPAL)}
\end{eqnarray}
and later by DELPHI Collaborations \cite{delphi}
based on the process  $e^{+}e^{-} \to e^{+}e^{-}\tau^{+}\tau^{-}$
\begin{eqnarray}
 -0.052 < a_{\tau} < 0.013 
\end{eqnarray}
It is clear that we need 
at least one order of magnitude  improvement 
to determine $a_{\tau}$. 
      
In the coupling of $\tau$ lepton to a photon,  
another interesting contribution is the  CP violating
effects which create electric dipol moment. CP violation 
has been observed  in the system of $K^{0}$ mesons \cite{cp1}.
This phenomenon has been described within the 
SM by the complex couplings 
in the Cabibbo-Kobayashi-Maskawa (CKM) matrix of the quark 
sector \cite{ckm}. Actually,  there is no CP violation 
in the leptonic couplings in the SM. In spite of that,
CP violation in the quark sector induces electric dipole moment
of the leptons in the three loop level \cite{hoog}. 
This contribution of the SM  to the 
electric dipole  moment of the leptons can be shown to be
too small to detect.    
Another source of CP violating coupling of leptons comes from 
the neutrino mixing if  neutrinos are massive \cite{barr}. It is also 
shown that this kind of CP violation is undetectable through 
the electric dipole moment of the $\tau$ lepton. 
Supersymmetry (SUSY) \cite{ellis} , more Higgs multiplets
\cite{weinberg}, left-right 
symmetric models \cite{pati} and leptoquarks \cite{ma}
 are expected to be 
the sources of the CP violation. 
Some loop diagrams are proportional to the  fermion masses 
which make the $\tau$  the most sensitive lepton to the 
CP violation. Therefore, larger effects may arise from the physics 
beyond the SM. Only upper limits on the electric 
dipole moment of the $\tau$ lepton
have been obtained so far from the experiments at 95\%CL 
\cite{l3,opal,delphi}
 
\begin{eqnarray}
|d_{\tau}| < 3.1\times 10^{-16}\,\, \mbox{e cm}\,\, \mbox{(L3)} \\
|d_{\tau}| < 3.7\times 10^{-16}\,\, \mbox{e cm}\,\, \mbox{(OPAL)} \\
|d_{\tau}| < 3.7\times 10^{-16}\,\, \mbox{e cm}\,\, \mbox{(DELPHI)}
\end{eqnarray}
More stringent limits were set by BELLE \cite{belle}  

\begin{eqnarray}
-0.22<Re(d_{\tau}) < 0.45 \,\,( 10^{-16}\,\, \mbox{e cm}) \\
-0.25<Im(d_{\tau}) < 0.08 \,\,( 10^{-16}\,\, \mbox{e cm}) 
\end{eqnarray}

There are more articles providing limits from 
previous LEP results \cite{cornet} or  obtained 
by using some indirect methods and early  study in 
 heavy ion collision  \cite{masso}.
   
Couplings of $\tau$ lepton to a photon can be parametrized  
by replacing the pointlike factor $\gamma^{\mu}$ 
by $\Gamma^{\mu}$ as follows \cite{grimus}

\begin{eqnarray}
 \Gamma^{\mu}=F_{1}(q^{2})\gamma^{\mu}+F_{2}(q^{2})
\frac{i}{2m_{\tau}}\sigma^{\mu\nu}q_{\nu}+F_{3}(q^{2})
\frac{1}{2m_{\tau}}\sigma^{\mu\nu}q_{\nu}\gamma^{5}
\end{eqnarray}
where  $F_{1}(q^{2})$, $F_{2}(q^{2})$ and $F_{3}(q^{2})$ 
are form factors related to electric charge, anomalous 
magnetic dipole moment and electric dipole moment. 
q is defined as the momentum transfer to the photon and 
$\sigma^{\mu\nu}=\frac{i}{2}(\gamma^{\mu}\gamma^{\nu}-
\gamma^{\nu}\gamma^{\mu})$.  
Asymptotic values of the form factors,
in the limiting case $q^{2}\to 0$,  are called moments 
describing  the static properties of the fermions
\begin{eqnarray}
F_{1}(0)=1,\,\,\, a_{\tau}=F_{2}(0), \,\,\, 
d_{\tau}=\frac{e}{2m_{\tau}}F_{3}(0)
\end{eqnarray}
 
In the next section, we give  some details of the equivalent 
photon approximation and forward detector physics at LHC. Then 
we study  the sensitivity of the process 
$pp \to pp \tau^{+}\tau^{-} $ to the anomalous 
electromagnetic  moments of the $\tau$ lepton via 
the subprocess $\gamma\gamma \to \tau^{+}\tau^{-} $.  

\section{$\gamma\gamma$ Scattering at LHC}

Two photon scattering physics at Large Hadron Collider (LHC)
is becoming interesting as an additional tool to search for 
physics in Standard Model (SM) or beyond it. Forward  
detectors  at ATLAS and CMS are developed to detect the particles 
not detected by the central detectors with a pseudorapidity
$\eta$ coverage 2.5 for tracking system and 5.0 for calorimetry. 
In many cases, the elastic scattering
and ultraperipheral collisions are out of the central detectors.
According to the program of ATLAS and CMS Collaborations forward 
detectors will be installed in a region nearly 100m-400m from the
interaction point \cite{royon}. With these new equipments, 
it is aimed to investigate soft and hard diffraction,
low-x dynamics with forward jet studies,
high energy photon induced interactions,
large rapidity gaps between forward jets,
and luminosity monitoring \cite{royon,khoze, schul}.
These dedicated detectors may tag protons with  
energy fraction loss $\xi =E_{loss}/E_{beam}$ far away from the 
interaction point. This nice property allows for 
high energy photon induced interactions with exclusive final states 
in the central detectors. 
In the recent program of ATLAS and CMS,  the positions of the 
forward detectors are planned to give an overall 
acceptance region of  $0.0015<\xi<0.5$ \cite{royon2,albrow}. 
Closer location of the forward detectors to interaction 
point leads to higher $\xi$. 
Almost real photons are emitted by each proton and  interact 
each other to produce exclusive final states.
In this work, we are interested
in the  $\tau$ lepton pair in the final states 
$\gamma\gamma \to \tau^{+}\tau^{-}$.
Deflected protons and their energy loss will be detected 
by the forward detectors far away from the 
interaction point as mentioned above.
Final $\tau$ leptons 
with rapidity $|\eta|<2.5$ and $p_{T}>20 GeV$ will be identified 
by the central detector. Photons emitted with small angles
by the protons show  a spectrum of virtuality $Q^{2}$ and energy 
$E_{\gamma}$. In order to handle  this kind of 
processes    equivalent 
photon approximation \cite{budnev,baur} is used.  
The proton-proton  case  differs from the pointlike 
electron-positron case by including the electromagnetic form 
factors in the equivalent photon spectrum and effective 
$\gamma\gamma$ luminosity

\begin{eqnarray}
dN=\frac{\alpha}{\pi}\frac{dE_{\gamma}}{E_{\gamma}}
\frac{dQ^{2}}{Q^{2}}[(1-\frac{E_{\gamma}}{E})
(1-\frac{Q^{2}_{min}}{Q^{2}})F_{E}+\frac{E^{2}_{\gamma}}{2E^{2}}F_{M}]
\end{eqnarray}
where

\begin{eqnarray}
Q^{2}_{min}=\frac{m^{2}_{p}E^{2}_{\gamma}}{E(E-E_{\gamma})}, 
\;\;\;\; F_{E}=\frac{4m^{2}_{p}G^{2}_{E}+Q^{2}G^{2}_{M}}
{4m^{2}_{p}+Q^{2}} \\
G^{2}_{E}=\frac{G^{2}_{M}}{\mu^{2}_{p}}=(1+\frac{Q^{2}}{Q^{2}_{0}})^{-4}, 
\;\;\; F_{M}=G^{2}_{M}, \;\;\; Q^{2}_{0}=0.71 \mbox{GeV}^{2}
\end{eqnarray}  
Here E is the energy of the proton beam which is related to the 
photon energy by $E_{\gamma}=\xi E$ 
and $m_{p}$ is the mass of the proton.
The magnetic moment of the proton is $\mu^{2}_{p}=7.78$, $F_{E}$ and 
$F_{M}$ are functions of the electric and magnetic form factors.
The integration of the subprocess $\gamma\gamma \to \tau^{+}\tau^{-}$
over the photon spectrum is needed 

\begin{eqnarray}
d\sigma=\int{\frac{dL^{\gamma\gamma}}{dW}} 
d\sigma_{\gamma\gamma \to \tau\tau}(W)dW
\end{eqnarray} 
where the effective photon luminosity $dL^{\gamma\gamma}/dW$ 
is given by  

\begin{eqnarray}
\frac{dL^{\gamma\gamma}}{dW}=\int_{Q^{2}_{1,min}}^{Q^{2}_{max}}
{dQ^{2}_{1}}\int_{Q^{2}_{2,min}}^{Q^{2}_{max}}{dQ^{2}_{2}}
\int_{y_{min}}^{y_{max}}
{dy \frac{W}{2y} f_{1}(\frac{W^{2}}{4y}, Q^{2}_{1}) 
f_{2}(y,Q^{2}_{2})}.
\end{eqnarray}
with 

\begin{eqnarray}
y_{min}=\mbox{MAX}(W^{2}/(4\xi_{max}E), \xi_{min}E), \;\;\;
y_{max}=\xi_{max}E, \;\;\;
f=\frac{dN}{dE_{\gamma}dQ^{2}}.
\end{eqnarray}
Here W is the invariant mass of the two photon system 
$W=2E\sqrt{\xi_{1}\xi_{2}}$ and $Q^{2}_{max}$ is the 
maximum virtuality.
Behaviour of the effective $\gamma\gamma$ luminosity
is shown in Fig.\ref{fig1} as a function of the invariant 
mass of the two photon system.
$Q^{2}_{max}$ dependence of the effective $\gamma\gamma$ luminosity 
will not be  separable in Fig.\ref{fig1} between 
$Q^{2}_{max}=(1-4)$ $\mbox{GeV}^{2}$. This is due to 
electromagnetic dipole form factors 
of the protons which  are steeply falling as a function of  $Q^{2}$. 
This causes very slow increase in $\gamma\gamma$ luminosity as 
$Q^{2}_{max}$ increases. This is explicitly shown in Table \ref{tab1} 
where the cross sections are calculated in the next section.
From Table \ref{tab1}, we see that $Q^{2}_{max}$ dependence 
does not create considerable uncertainty.
Thus, it is reasonable to take  
$Q^{2}_{max}$ as (1-2)$\mbox{GeV}^{2}$.  

\FIGURE{\epsfig{file=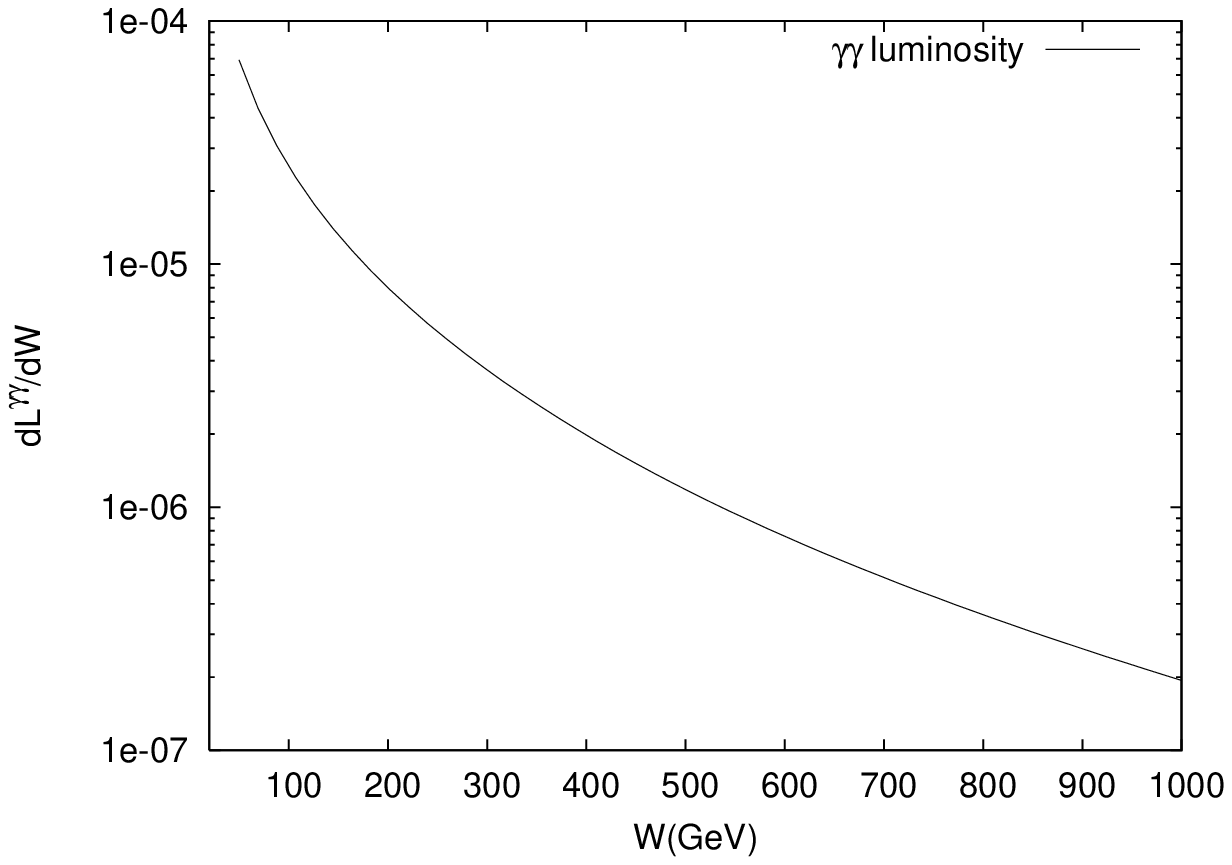}
\caption{Effective $\gamma\gamma$ luminosity as a function
of the invariant mass of the two photon system.}
\label{fig1}}
  
\TABLE{
\begin{tabular}{|c|c|c|}\hline
$Q^{2}_{max}(GeV^{2})$ & $\sigma^{0}$(fb) & $\sigma^{0}$(fb) \\
\hline
     & $ 0.0015<\xi<0.5$ & $ 0.01<\xi<0.15$  \\
\hline
0.5 & 167.6 & 10.4 \\
0.8 & 171.3 & 10.7 \\
1 & 172.3 & 10.8 \\
1.5 & 173.3 & 10.9 \\
1.8 & 173.5 & 10.9 \\
2 & 173.6 & 10.9 \\
3 & 173.8 & 11.0 \\
4 & 173.8 & 11.0 \\ 
\hline
\end{tabular}
\caption{$Q^{2}_{max}$ dependence of the cross sections 
with equivalent photon approximation   
for the process $ pp \to p\tau^{+}\tau^{-}p$
without anomalous couplings of tau lepton . 
Two intervals of forward detector acceptance $\xi$ 
are considered. For $Q^{2}_{max}=(1-4)$ $GeV^{2}$  
the cross sections do not change appreciably.   
\label{tab1}}}   

There are experimental uncertainties in the dipole form 
factors in Eq. (2.3). In Ref. \cite{arington} these 
uncertainties are given for the region $Q^{2}=0.007-5.850$ $GeV^{2}$. 
The change in the photon flux $f(E_{\gamma}, Q^{2})$ from 
the uncertainties in the electric and magnetic form factors
can be calculated with the help of  the expression below

\begin{eqnarray}
\delta f=\sqrt{(\frac{\partial{f}}{\partial{G_{E}}}\delta G_{E})^2
 +(\frac{\partial{f}}{\partial{G_{M}}}\delta G_{M})^2}
\end{eqnarray}
Using some of the   uncertainties in Ref. \cite{arington} we obtain 
relative changes in the photon flux $\delta f/f$. The results 
are shown in  Table \ref{tab2} for two  photon energies.
The uncertainty in the photon flux from both protons 
leads to the relative
uncertainty   in the cross section
$\delta\sigma/\sigma $ around 0.03 on the average 
depending on the photon energy    for the
process $ pp \to p\tau^{+}\tau^{-}p$ with 
$Q^{2}_{max}=2$ $GeV^{2}$.

\TABLE{
\begin{tabular}{|c|c|c|c|c|}\hline
$\xi$ &$Q^{2}(GeV^{2})$ & $\delta G_{E}/G_{D}$ & 
$\delta G_{M}/(\mu_{p}G_{D})$ & $\delta f/f$ \\
\hline
0.01 & 0.022 & 0.003 & 0.019 & 0.006   \\
0.01 & 0.115 & 0.011 & 0.007 & 0.018  \\
0.01 & 0.528 & 0.013 & 0.009 & 0.015  \\
0.01 & 1.020 & 0.017 & 0.006 & 0.013  \\
0.01 & 2.070 & 0.038 & 0.006 & 0.017  \\
\hline
0.15 & 0.022 & 0.003 & 0.019 & 0.090  \\
0.15 & 0.115 & 0.011 & 0.007 & 0.016  \\
0.15 & 0.528 & 0.013 & 0.009 & 0.015  \\
0.15 & 1.020 & 0.017 & 0.006 & 0.013  \\
0.15 & 2.070 & 0.038 & 0.006 & 0.017  \\
\hline
\end{tabular}
\caption{Relative change in the photon flux 
$\delta f/f$  due to the
experimental uncertainties in dipole form factors. 
Values in the middle three columns are taken from 
Ref. \cite{arington}.
\label{tab2}}}

Let us discuss briefly bremsstrahlung lepton pair production 
which is one of the possible backgrounds to the equivalent photon 
approximation. In this process, there are a virtual photon exchange 
between the two protons and one bremsstrahlung photon  emitted 
by one of the protons. The bremsstrahlung photon creates 
a lepton pair.  The square of the matrix element includes 
electromagnetic form factors in each of the photon-proton vertex
which are given in Ref.\cite{baur2} 
\begin{eqnarray}
|M_{if}|^{2} \to |M_{if}|^{2} |F_{A}(q^{2}_{1})|^{2} 
|F_{B}(q^{2}_{1})|^{2}|F_{T}(q^{2}_{2})|^{2}
\end{eqnarray}
where $q_{1}$ is the momentum transfer between two protons and 
$q_{2}$ is identical to the momentum of the lepton pair. 
$F_{A}(q^{2}_{1})$, $F_{B}(q^{2}_{1})$ are elastic form 
factors in the space-like region and 
$F_{T}(q^{2}_{2})$ is the form factor 
in the time-like region.
If we have high $q_{2}^{2}$ the form factor 
$|F_{T}(q^{2}_{2})|^{2}$ 
will supress the cross section based on the fact that the 
large $q^{2}$ form factors behave like $1/q^{4}$. 
In our work, as will be seen 
in the next section, each tau lepton 
in the final state has $p_{T}>20$ GeV. Therefore the minimum 
$q_{2}^{2}$ value is $4(m_{\tau}^{2}+p_{T}^{2})=1612$ 
$\mbox{GeV}^{2}$ which makes the cross section for the  
bremsstrahlung tau pair production completely negligible.

Two photon exchange interactions with
invariant diphoton mass $W > 1$ TeV are highly interesting to 
probe more accurate values of the SM parameters and  also
deviations from SM with available luminosity. 

\section{Cross Sections And Sensitivity}

There are t and u channels Feynman diagrams of the 
subprocess $\gamma\gamma \to \tau^{+}\tau^{-}$ where 
both vertices contain anomalous couplings.   
The squared amplitude can be written in terms of the 
following reduced amplitudes,

\begin{eqnarray}
A_{1}&&=\frac{1}{2m^{4}}[48F_{1}^{3}F_{2}(m^{2}-\hat{t})
(m^{2}+\hat{s}-\hat{t})m^{4}-16F_{1}^{4}(3m^{4}-\hat{s}m^{2}
+\hat{t}(\hat{s}+\hat{t}))m^{4} \nonumber \\
&&+2F_{1}^{2}(m^{2}-\hat{t})
(F_{2}^{2}(17m^{4}+(22\hat{s}-26\hat{t})m^{2}+\hat{t}
(9\hat{t}-4\hat{s})) \nonumber \\
&&+F_{3}^{2}
(17m^{2}+4\hat{s}-9\hat{t})(m^{2}-\hat{t}))m^{2} \nonumber \\ 
&&+12F_{1}F_{2}(F_{2}^{2}+F_{3}^{2})\hat{s}(m^{3}-m\hat{t})^{2}
-(F_{2}^{2}+F_{3}^{2})^{2}(m^{2}-\hat{t})^{3}
(m^{2}-\hat{s}-\hat{t})]
\end{eqnarray}

\begin{eqnarray}
A_{2}&&=-\frac{1}{2m^{4}}[48F_{1}^{3}F_{2}(m^{4}+
(\hat{s}-2\hat{t})m^{2}+\hat{t}(\hat{s}+\hat{t}))m^{4}\nonumber \\
&&+16F_{1}^{4}(7m^{4}-(3\hat{s}+4\hat{t})m^{2}+
\hat{t}(\hat{s}+\hat{t}))m^{4} \nonumber \\
&&+2F_{1}^{2}(m^{2}-\hat{t})
(F_{2}^{2}(m^{4}+(17\hat{s}-10\hat{t})m^{2}
+9\hat{t}(\hat{s}+\hat{t})) \nonumber \\
&&+F_{3}^{2}(m^{2}-9\hat{t})(m^{2}-\hat{t}-\hat{s}))
m^{2}+(F_{2}^{2}+F_{3}^{2})^{2}(m^{2}-\hat{t})^{3}
(m^{2}-\hat{s}-\hat{t})]
\end{eqnarray}

\begin{eqnarray}
A_{12}&&=\frac{1}{m^{2}}[-16F_{1}^{4}(4m^{6}-m^{4}\hat{s})
+8F_{1}^{3}F_{2}m^{2}(6m^{4}-6m^{2}(\hat{s}+2\hat{t})-\hat{s})^{2}
+6\hat{t})^{2}+6\hat{s}\hat{t}) \nonumber \\
&&+F_{1}^{2}(F_{2}^{2}(16m^{6}-m^{4}(15\hat{s}+32\hat{t})+
m^{2}(-15\hat{s})^{2}+14\hat{t}\hat{s}+16\hat{t})^{2})
+\hat{s}\hat{t}(\hat{s}+\hat{t})) \nonumber \\
&&+F_{3}^{2}(16m^{6}-
m^{4}(15\hat{s}+32\hat{t})+m^{2}(-5\hat{s})^{2}+14\hat{t}\hat{s}
+16\hat{t})^{2})+\hat{s}\hat{t}(\hat{s}+\hat{t}))) \nonumber \\
&&-4F_{1}F_{2}(F_{2}^{2}+F_{3}^{2})\hat{s}(m^{4}+
m^{2}(\hat{s}-2\hat{t})+\hat{t}(\hat{s}+\hat{t}))\nonumber \\
&&-4F_{1}F_{3}(F_{2}^{2}+F_{3}^{2})(2m^{2}-\hat{s}-2\hat{t})
\epsilon_{\mu\nu\rho\sigma}p_{1}^{\mu}p_{2}^{\nu}p_{3}^{\rho}
p_{4}^{\sigma} \nonumber \\
&&-2(F_{2}^{2}+F_{3}^{2})^{2}\hat{s}
(m^{4}-2\hat{t}m^{2}+\hat{t}(\hat{s}+\hat{t}))]
\end{eqnarray}
where $p_{1}$, $p_{2}$, $p_{3}$ and $p_{4}$  are the 
momenta of the incoming photons and final $\tau$ leptons.
Mandelstam variables are defined as $\hat{s}=(p_{1}+p_{2})^{2}$,
$\hat{t}=(p_{1}-p_{3})^{2}$ and $\hat{u}=(p_{1}-p_{4})^{2}$. 
m is the $\tau$ lepton mass. The squared amplitudes are 

\begin{eqnarray}
|M_{1}|^{2}&&=\frac{16\pi^{2}\alpha^{2}}{(\hat{t}-m^{2})^{2}}A_{1} \\
|M_{2}|^{2}&&=\frac{16\pi^{2}\alpha^{2}}{(\hat{u}-m^{2})^{2}}A_{2} \\
|M_{12}|^{2}&&=\frac{16\pi^{2}\alpha^{2}}{(\hat{t}-m^{2})
(\hat{u}-m^{2})}A_{12} 
\end{eqnarray}

The cross section  for the process $pp \to pp\tau^{+}\tau^{-}$ 
without anomalous couplings is given in Table \ref{tab3} 
at the LHC energy $\sqrt{s}=14$ TeV 
for rapidity $\eta<2.5$ and transverse 
momentum $p_{T}>20$ GeV of the final $\tau$ leptons. 

The possible  background is the diffractive double pomeron exchange
(DPE) production of tau pairs created via  Drell-Yan process.
The DPE production cross section can be obtained  within 
the factorized Ingelman-Schlein \cite{ingelman}
 model where the concept 
of diffractive parton distribution function(DPDF) is introduced.
The convolution integral for the subprocess 
$ q\bar{q}\to \tau\tau$ is given by   

\begin{eqnarray}
\sigma&&=\int dx_{1} dx_{2} d\beta_{1} d\beta_{2}
f_{\mathbb{P}/p}(x_{1},t) f_{\mathbb{P}/p}(x_{2},t) \nonumber  \\  
&&\sum_{i,j=1}^{3} \left [f_{i}(\beta_{1},Q^{2}) f_{j}(\beta_{2},Q^{2})+
f_{j}(\beta_{1},Q^{2}) f_{i}(\beta_{2},Q^{2})\right ]
\hat{\sigma}(q\bar{q}\to \tau\tau)
\end{eqnarray}
where $f_{\mathbb{P}/p}(x_{1},t)$ is the pomeron flux emitted by 
one of the protons and $f_{i}(\beta_{1},Q^{2})$ is the light quark
distribution function coming from the structure of the pomeron.
$x_{1}$, $x_{2}$ denote the momentum fractions of the protons 
carried by the 
pomeron fluxes and $\beta_{1}$, $\beta_{2}$ represent the 
longitudinal momentum fractions of the pomeron  
carried by the struck quarks.  
Double pomeron exchange production cross section should be multiplied 
by gap survival probability 0.03 for LHC.
The measurements of pomeron flux and DPDF were performed at HERA 
with their uncertainties \cite{aktas,aktas2}.
The uncertainty in DPDF was obtained as (5-10)\% for light quarks
in Fig.11 of Ref. \cite{aktas2}. We have 
determined  the uncertainty in the  pomeron flux as (8-10)\%
using the uncertainties of the flux  parameters which 
were given in  Ref. \cite{aktas}. Taking the maximum values 
of the each uncertainties above,   the combined uncertainty 
due to both  DPDF  and pomeron flux from one proton 
is estimated by 14\%. 
The overall uncertainty related to pomerons arising from both 
protons is expected to be  20\% using a root sum-of-the-squares 
approach.  
\TABLE{
\begin{tabular}{|c|c|c|}\hline
$\xi$ &$\sigma^{\mathbb{P}}$ (fb) & $\sigma^{0}$ (fb)\\
\hline
0.0015-0.5 & 28.4$\pm$ 2.8   & 173$\pm$ 2.6   \\
0.0015-0.15 & 27.2$\pm$ 2.7  & 173$\pm$ 2.6  \\
0.01-0.15 &  4.6$\pm$ 0.5 &  10.9$\pm$ 0.2  \\
\hline
\end{tabular}
\caption{Cross sections    $\sigma^{\mathbb{P}}$
obtained by double  pomeron exchange
production of tau pairs  multiplied by
gap survival probability 0.03.
For comparison, the cross sections $\sigma^{0}$
for the same subprocess   obtained by equivalent
photon approximation at tree level
(without anomalous couplings)  are given.
In both cases the LHC energy $\sqrt{s}=14$ TeV,
transverse momentum and rapidity cuts
$p_{T}>20$ $GeV$ and $|\eta|< 2.5$ are taken
into account for each final $\tau$ lepton.
The uncertainty in the $\sigma^{\mathbb{P}}$ is due to
pomeron flux and DPDF. The uncertainty in the $\sigma^{0}$
is related to the dipole form factors in the equivalent
photon spectrum.   \label{tab3}}}
Considering $t=-1$ $GeV^{2}$, 
$Q^{2}=2$ $GeV^{2}$ the calculated cross sections are given 
in Table \ref{tab3} for three acceptance regions of 
forward detectors. During computation sum over 
three light quarks  in Eq. (3.7) has been  considered.
Measurements at HERA for pomeron flux and DPDF have 
ranges  $8.5<Q^{2}<1600$ $GeV^{2}$ and $0.08<|t|<0.5$. Using 
NLO DGLAP equations DPDF were evolved to higher and lower 
scales, beyond the measured range,
and the grids were provided  for  $1<Q^{2}<30000$ $GeV^{2}$
in the H1 2006 DPDF Fits. The data were also analysed 
by integrating the cross section over the range 
$t_{min}<|t|<1$ $GeV^{2}$ \cite{aktas}. 
Possible additional uncertainties from these extrapolations
are expected to be compansated by choosing maximum 
individual uncertainties before combination.
Anomalous couplings are more sensitive to higher energies 
based on the term $\sigma_{\mu\nu}q^{\nu}$. For the invariant 
two photon mass $W>1$ TeV with sufficent luminosity
we are expecting far better result than 
the case of LEP energies. First we place bounds 
on the tau  anomalous magnetic moment  
by $\chi^{2}$ analysis  keeping $F_{3}=0$. 
  
\begin{eqnarray}
\chi^{2}=\frac{(\sigma(F_{2})-\sigma^{0})^{2}}
{(\sigma^{0}+\sigma^{\mathbb{P}})^{2}\delta^{2}}\\
\delta=\sqrt{(\delta^{st})^{2}+(\delta^{sys})^{2}} \\
\delta^{st}=\frac{1}{\sqrt{N^{0}}}\\
N^{0}=L_{int}(\sigma^{0}+\sigma^{\mathbb{P}}) BR
\end{eqnarray}
where $\sigma^{0}$, $N^{0}$ and $\delta$ are the cross section, 
number of events and uncertainty without anomalous couplings.
$L_{int}$ is the integrated luminosity of LHC.
The contributions of pomeron background do not appear 
in the numerator because of  cancellation of each other.
Thus, the pomeron contribution
in the denominator is not expected to be  so effective 
even if it has large 20\% uncertainty.
Now let us determine the effect of uncertainties due to 
$\sigma(F_{2})$, $\sigma^{0}$ and pomeron backgrounds on the 
$\chi^{2}$ function. The sources of uncertainties of 
$\sigma(F_{2})$ and $\sigma^{0}$ are connected to the dipole 
form factors in the equivalent photon spectrum, as explained before.
The change in the $\chi^{2}$ function from the 
3\% uncertainty of $\sigma(F_{2})$ and $\sigma^{0}$
lead to the $\delta^{sys}$ values shown in Table \ref{tab4}.
Total systematic uncertainty can be formed 
by combining individual contributions in quadrature  given 
in the last column of the Table \ref{tab4} . 
In our calculations, the individual 
uncertainties have been kept maximum and have been 
considered to be uncorrelated to get larger systematic uncertainty
$\delta^{sys}$.

\TABLE{
\begin{tabular}{|c|c|c|c|c|c|}\hline
$L_{int}(fb^{-1})$ & $\xi$ &$\delta^{sys}(\sigma(F_{2,3}))$ &
 $\delta^{sys}(\sigma^{0})$ & $\delta^{sys}(\sigma^{\mathbb{P}})$
& $\delta^{sys}$ \\
\hline
50  & 0.0015-0.5&0.016&  0.016 & 0.002 & 0.02  \\
100 & 0.0015-0.5&0.014 & 0.014 & $<$0.002 &  0.02\\
200 & 0.0015-0.5 & 0.012& 0.012 & $<$0.002 & $<$0.02 \\
\hline
50  & 0.0015-0.15&0.016&  0.016 & 0.002 & 0.02  \\
100 & 0.0015-0.15&0.014 & 0.014 & $<$0.002 &  0.02\\
200 & 0.0015-0.15 & 0.012& 0.012 & $<$0.002 & $<$0.02 \\
\hline
50  & 0.01-0.15&0.025  &  0.025 & 0.007 & 0.04  \\
100 & 0.01-0.15&0.020 &   0.020 & 0.006 &  0.03\\
200 & 0.01-0.15 & 0.018 & 0.018 & 0.005 & $<$0.03 \\
\hline
\end{tabular}
\caption{Systematic uncertainties in the 
$\chi^2$ function  depending on 
luminosity and acceptance region $\xi$. The 
last column represents the combined uncertainties
in quadrature. The values with $<$ character defines 
the uncertainties less than the specified values. 
\label{tab4}}}

In this work all computations are done in the laboratory
frame of the two protons. 
For the signal we consider one of  the tau leptons 
decays hadronically and 
the other leptonically with branching ratios 65\% and 35\%. 
Then joint branching ratio of the tau pairs becomes 
BR=0.46. 

\TABLE{
\begin{tabular}{|c|c|c|c|}\hline
$L_{int}(fb^{-1})$ & $\xi$ & $a_{\tau}$ & $|d_{\tau}|$(e cm) \\ 
\hline
50  & 0.0015-0.5&-0.0062, 0.0042 &0.23$\times 10^{-16}$ \\
100 & 0.0015-0.5&-0.0057, 0.0037 & 0.21$\times 10^{-16}$ \\ 
200 & 0.0015-0.5 &-0.0054, 0.0034 & 0.19$\times 10^{-16}$ \\
\hline
50  & 0.0015-0.15&-0.0063, 0.0043 &0.23 $\times 10^{-16}$ \\
100 & 0.0015-0.15&-0.0058, 0.0037 & 0.22 $\times 10^{-16}$ \\
200 & 0.0015-0.15 &-0.0055, 0.0034 & 0.20 $\times 10^{-16}$ \\
\hline
50  & 0.01-0.15&-0.0048, 0.0045 &0.19 $\times 10^{-16}$ \\
100 & 0.01-0.15&-0.0042, 0.0038 & 0.16 $\times 10^{-16}$ \\
200 & 0.01-0.15 &-0.0036, 0.0032 & 0.14 $\times 10^{-16}$ \\
\hline
\end{tabular} 
\caption{Sensitivity of the process
$pp\to p \tau^{+}\tau^{-}p$  to tau anomalous
 magnetic moment $a_{\tau}$ and electric
dipole moment $d_{\tau}$  at 95\% C.L. for
$\sqrt{s}=14$ TeV, integrated luminosities
$L_{int}=50,\,\,100,\,\,200$ $fb^{-1}$ and three
intervals of forward detector acceptance $\xi$.
Only one of the moments is assumed to deviate
from  zero at a time. Total systematic uncertainty used in 
$\chi^{2}$ function has been taken $\delta^{sys}=0.01$.
\label{tab5}}}

\TABLE{
\begin{tabular}{|c|c|c|c|c|}\hline
$L_{int}(fb^{-1})$ & $\xi$ &$\delta^{sys}$ &
 $a_{\tau}$ & $|d_{\tau}|$(e cm) \\
\hline
50  & 0.0015-0.5&0.02 &-0.0071, 0.0051 &0.28$\times 10^{-16}$ \\
100 & 0.0015-0.5&0.02 &-0.0068, 0.0048 & 0.26$\times 10^{-16}$ \\
200 & 0.0015-0.5 & 0.02& -0.0066, 0.0046 & 0.26$\times 10^{-16}$ \\
\hline
50  & 0.0015-0.15& 0.02 &-0.0073, 0.0051 &0.28 $\times 10^{-16}$ \\
100 & 0.0015-0.15&0.02  &-0.0070, 0.0048 & 0.27 $\times 10^{-16}$ \\
200 & 0.0015-0.15 &0.02  &-0.0067, 0.0048 & 0.27 $\times 10^{-16}$ \\
\hline
50  & 0.01-0.15&0.04  &-0.0054, 0.0050 &0.21 $\times 10^{-16}$ \\
100 & 0.01-0.15&0.03  &-0.0046, 0.0042 & 0.18 $\times 10^{-16}$ \\
200 & 0.01-0.15 &0.03  & -0.0043, 0.0038 & 0.17 $\times 10^{-16}$ \\
\hline
\end{tabular}
\caption{The same as the Table\ref{tab5} but for 
the systematic uncertainties shown in the third 
column.  \label{tab6}}}

Table \ref{tab5}  and Table \ref{tab6} show the constraints 
on the anomalous magnetic moment
and electric dipole moment of the tau lepton that we obtain 
using different systematic uncertainties in 
$\chi^{2}$ function for comparison.
The acceptance region $\xi=0.01-0.15$ seems more sensitive 
to anomalous couplings.
The limits are improved by one order of magnitude 
when compared to DELPHI 
results. Electric dipole moment limits are slightly better than 
those of BELLE. At this point a remark is in order. Experimentally, 
the anomalous magnetic and electric dipole moments can be extracted 
by comparing the measured cross section with QED expectations. 
At LEP \cite{delphi}, for example, 
the fits to the measured cross section were performed taking 
$a_{\tau}$ and $d_{\tau}$ as parameters based on 
the $\tau\tau\gamma$  vertex parametrization given by (1.12).
However, our predictions for the cross sections in the 
$\chi^{2}$ function are theoretical. When comparing our limits 
with those of  LEP this distinction should be taken into account. 
The quadratic and quartic terms according to 
$F_{3}$ are not CP violating except the term with Levi-Civita tensor 
in the interference amplitude $A_{12}$. However its contribution 
to the cross section is zero. That is why the 
magnitudes of negative and positive parts 
of the limits on  $d_{\tau}$ are the same.
This leads to the fact that it may be possible 
to measure tau anomalous magnetic moment when efficient tau 
identification is available. 

Tau is the heaviest charged lepton which decays 
into lighter leptons, electron, muon  and lighter hadrons such as 
$\pi$'s and $K$'s  with a lifetime of $3.0 \times 10^{-13}$ s. 
Primary decay channels can be given  
with one charged particle (one prong decay)  

\begin{eqnarray}
&&\tau \to \nu_{\tau}+\ell+\hat{\nu}_{\ell}, \,\,\,\,\ell=e, \mu \\
&&\tau \to \nu_{\tau}+ \pi^{\pm} \\
&&\tau \to \nu_{\tau}+ \pi^{\pm}+ \pi^{0} \\
&&\tau \to \nu_{\tau}+ \pi^{\pm}+ \pi^{0}+\pi^{0}
\end{eqnarray}  
and with three charged particle (three prong decay)
\begin{eqnarray}
\tau \to \nu_{\tau}+ 3\pi^{\pm}+n\pi^{0} 
\end{eqnarray}
85\% of all the tau decays are the one prong decays and 15\% of them
are the three prong decays.  
Produced particles from tau decays are called tau jets due to 
the fact that number of daughter particles is  always 
greater than one. One prong lepton jets are identified by similar 
algorithms used by direct electron and muon. Identification 
of hadronic jets is more complicated than leptonic modes because 
of the QCD jets as background. However, tau jets are higly collimated 
and are distingushed from background due to its topology.  
Dedicated algorithms have been developed for hadronic tau jets by 
ATLAS \cite{atlas} and CMS \cite{cms} groups. 
Use of these algorithms allows for good
separation between  tau jets and  fake jets for some LHC process. 
Nevertheless, tau identification efficiency depends of a specific 
process, background processes, some kinematic parameters and 
luminosity. Studies of tau identification have not been finalized 
yet for  LHC detectors.
In every case, identification efficiency can be determined 
as a function of transverse momentum and rapidity. 
In our study we have considered 
$p_{T}>20$ GeV and $\eta<2.5$ for a good $\tau$ selection
as used in most  ATLAS and CMS studies.
For a realistic efficiency we need a detailed study based on 
our specific process including properties of  both central and    
forward detectors of  ATLAS and CMS experiments. We expect 
highly efficient $\tau$ identification due to clean final state 
in the process $\gamma\gamma \to \tau^{+}\tau^{-}$ 
when compared to the LHC itself.

\end{document}